\documentclass[11pt]{article}
\usepackage[utf8]{inputenc}
\usepackage[T1]{fontenc}
\usepackage{lmodern}
\usepackage[margin=2.7cm]{geometry}
\usepackage{amsmath,amssymb,amsthm}
\usepackage[linesnumbered,ruled,vlined]{algorithm2e}
\usepackage{enumitem}
\usepackage{booktabs}
\usepackage{tikz}
\usepackage[hidelinks]{hyperref}
\usepackage[capitalise]{cleveref}

\newtheorem{theorem}{Theorem}[section]
\newtheorem{lemma}[theorem]{Lemma}

\newtheorem{corollary}[theorem]{Corollary}
\theoremstyle{definition}

\newtheorem{example}[theorem]{Example}
\theoremstyle{remark}

\crefname{algocf}{Algorithm}{Algorithms}
\Crefname{algocf}{Algorithm}{Algorithms}

\newcommand{\E}{\mathbb{E}}
\newcommand{\Prob}{\mathbb{P}}

\newcommand{\Ent}{\mathcal{H}}
\newcommand{\W}{\mathcal{W}}

\renewcommand{\lg}{\log_2}
\SetKwFunction{FDR}{Uniform}
\SetKwFunction{Insert}{Insert}

\title{First-Order Entropic Sampling of Words with Prescribed Letter Counts in Linear Time}
\author{Olivier Bodini\thanks{LAGA, Institut Galilée, Université Sorbonne Paris Nord, CNRS, UMR 7539, F-93430 Villetaneuse, France. \texttt{olivier.bodini@univ-paris13.fr}}
\and Francis Durand\thanks{LAGA, Institut Galilée, Université Sorbonne Paris Nord, CNRS, UMR 7539, F-93430 Villetaneuse, France. \texttt{francis.durand@ens-paris-saclay.fr}}}
\date{}

\begin{document}
\maketitle

\begin{abstract}
We sample, exactly and uniformly, a word of length $n$ over the alphabet $\{0,\dots,r\}$ in which the letter $j$ occurs a prescribed positive number $k_j$ of times, from a source of fair random bits.
Every exact sampler reads on average at least $\Ent=\lg\binom{n}{k_0,\dots,k_r}$ bits.
We give a sampler that reads on average at most $\big(1+O(1/\log n)\big)\Ent$ fresh bits, whatever $r$ and the counts, and that performs $O(n)$ bit operations on average over every fixed alphabet, in a bit-RAM in which a memory access costs $O(1)$ while arithmetic is charged bit by bit.
Both bounds are optimal: the first one to first order, the second one up to a constant factor, since writing the output already costs $\Theta(n)$.
The sampler writes the word as a Huffman tree of independent binary words, and samples each of them in one of two ways.
Binary words with few ones are generated by random insertions; the rank of each inserted letter among its copies, which the Fisher--Yates shuffle generates for nothing, is uniform and independent of the word, and is given back to a recycling source of uniform integers, where it pays for the next insertions.
Binary words with many ones are generated by \textsc{MergeShuffle} driven by blocks of $\Theta(\log n)$ bits of entropy; a block is decoded by locating only its ones when the minority letter is rare, and otherwise, its number of ones being concentrated, by a precomputed table of the blocks of typical type, of size $O(\sqrt n\log^4n)$, the rare atypical blocks being decoded slowly.
\end{abstract}

\section{Introduction}\label{sec:intro}

Let $k=(k_0,\dots,k_r)$ be positive integers and $n=k_0+\dots+k_r$.
We write $\W(k)$ for the set of words of length $n$ over the alphabet $\{0,\dots,r\}$ in which the letter $j$ occurs exactly $k_j$ times, and
\[
\Ent=\lg|\W(k)|=\lg\binom{n}{k_0,\dots,k_r}.
\]
For $r=1$, this is the sampling of a $k_1$-subset of an $n$-set.
We want an exact sampler for the uniform distribution on $\W(k)$ that uses a source of fair random bits, and we care about two costs: the expected number of random bits that it reads, and its expected running time.
Every exact sampler reads on average at least $\Ent$ bits~\cite{KnuthYao1976}, and we call a family of samplers \emph{first-order entropic} if it reads $\Ent+o(\Ent)$ bits on average.
We measure time by the number of bit operations in a \emph{bit-RAM} (\cref{sec:model}), in which reading or writing a bit of memory at an address held in a register costs $O(1)$, while arithmetic is charged bit by bit.
Since writing the output word already costs $\Theta(n)$ in this model over a fixed alphabet, $O(n)$ bit operations is the best one can hope for.

None of the classical samplers achieves both bounds.
Enumerative decoding~\cite{Cover1973} of a uniform integer drawn by the Fast Dice Roller~\cite{Lumbroso2013} reads at most $\Ent+2$ bits, but it relies on multiprecision arithmetic on integers of about $\Ent$ bits; even with the fast algorithms of~\cite{Ryabko1998,HarveyVanDerHoeven2021}, the best bound that we know for it in our model is a polylogarithmic factor above $n$.
At the other extreme, the partial Fisher--Yates shuffle~\cite{Durstenfeld1964}, which inserts the non-majority letters one by one at uniform positions, is fast but wasteful: it reads about $\Ent+\sum_{j\ne0}\lg k_j!$ bits, because it also generates the order in which the copies of each letter are inserted.
\textsc{MergeShuffle}~\cite{BacherBodiniHollenderLumbroso2015}, which draws independent letters until one of them overflows and completes the word by insertions, lies in between: it is close to entropic when all the counts are large, but performs too many insertions otherwise.
Our main result shows that both bounds can be achieved at once.

\begin{theorem}[informal version of \cref{thm:main}]\label{thm:intro}
There is a sampler for the uniform distribution on $\W(k_0,\dots,k_r)$ that reads on average at most $\big(1+O(1/\log n)\big)\Ent$ fresh bits, uniformly in $r$ and in the positive counts, and performs $O\big(n(1+\lg(r+1))^2+r\sqrt n\big)$ bit operations on average; in particular, $O(n)$ for every fixed $r$.
\end{theorem}

The sampler reduces the problem to binary words, by writing the word as a Huffman tree of independent binary words.
All the uniform integers are drawn from a recycling source, an extension of the Fast Dice Roller of Lumbroso~\cite{Lumbroso2013} that keeps the unused part of its draws as a uniform integer and can take back uniform integers produced by the algorithm.
A binary word with few ones is generated by the Fisher--Yates shuffle, but the rank of each inserted one among the ones, which the shuffle generates for nothing and which is uniform and independent of the new word, is given back to the source, where it pays for the next insertions; the idea of giving these bits back comes from the preliminary version~\cite{BodiniDurandGascom}.
A binary word with many ones is generated by \textsc{MergeShuffle}, whose letters are drawn by blocks, and this is where the main difficulty lies.
A block of entropy $B$ wastes up to $2$ bits, the overhead of the Knuth--Yao sampler that draws its number of ones, so that $B$ must tend to infinity for the sampler to be first-order entropic; but decoding a block by the usual unranking handles integers of about $B$ bits at every position, and costs $\Theta(B)$ bit operations per letter.
We take $B$ of order $\lg n$, and decode a block in $O(1)$ bit operations per letter on average.
When the minority letter is rare, a block has few ones, and we locate only them.
Otherwise, the number of ones of a block is concentrated around its mean, and a precomputed table of the blocks of typical type, of size $O(\sqrt n\log^4n)$, decodes all the blocks but a fraction $O(\log^{-3}n)$ of them, which are decoded slowly.
The excess $O(1/B)$ of random bits per bit of entropy is responsible for the rate $1/\log n$.

The whole sampler is described in \cref{sec:algorithm}, which can be read on its own and ends with the statement of the main theorem.
After the model and the classical tools on random bits (\cref{sec:model}), \cref{sec:sparse,sec:dense} analyse the two samplers of binary words, and \cref{sec:proof} puts everything together.
We conclude in \cref{sec:conclusion} with a discussion of the rate and some open questions.

\section{The sampler}\label{sec:algorithm}

The first step is to replace the word by binary words.
Let $T$ be a full binary tree whose leaves are the letters $0,\dots,r$.
For an internal node $v$, let $s_v$ be the total count of the leaves below $v$, let $m(v)$ be a child of $v$ of smaller total count (ties being broken deterministically), and let $a_v\le s_v/2$ be the total count below $m(v)$; since the counts are positive, $a_v\ge1$.
Given a word $w\in\W(k)$, let $w_{|v}$ be the subsequence of $w$ formed by the letters below $v$; the \emph{node word} $B_v(w)\in\{0,1\}^{s_v}$ has a $1$ at position $i$ exactly when the $i$-th letter of $w_{|v}$ lies below $m(v)$.
In other words, $B_v(w)$ records, for each letter below $v$, the subtree in which it goes, and it belongs to $\W(s_v-a_v,a_v)$.

\begin{figure}[htbp]
\centering
\begin{tikzpicture}[level distance=13mm,
  every node/.style={font=\small},
  level 1/.style={sibling distance=40mm},
  level 2/.style={sibling distance=22mm}]
\node[draw,rounded corners,align=center] {root: $s=8$, $a=4$\\ $B=01101001$}
  child {node[draw,circle] {$\mathtt a$} edge from parent node[left] {\scriptsize $0$}}
  child {node[draw,rounded corners,align=center] {$u$: $s=4$, $a=2$\\ $B=1001$}
    child {node[draw,circle] {$\mathtt b$} edge from parent node[left] {\scriptsize $0$}}
    child {node[draw,circle] {$\mathtt c$} edge from parent node[right] {\scriptsize $1$}}
    edge from parent node[right] {\scriptsize $1$}};
\end{tikzpicture}
\caption{The tree of \cref{ex:tree} and the node words of $w=\mathtt{acbabaac}$; the edge labelled $1$ leads to the minority child.}\label{fig:tree}
\end{figure}
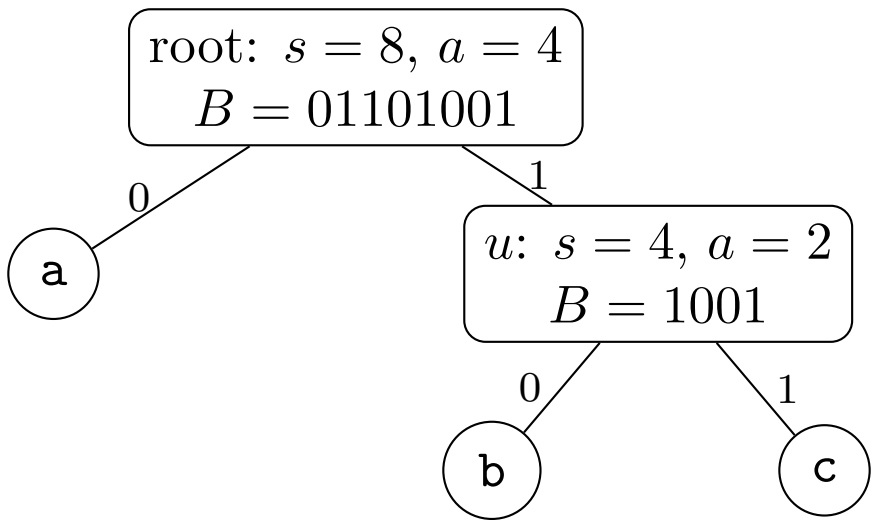

\begin{example}\label{ex:tree}
Take $k=(4,2,2)$ over the letters $\mathtt a,\mathtt b,\mathtt c$, and the tree of \cref{fig:tree}: an internal node $u$ above $\mathtt b$ and $\mathtt c$, with $m(u)=\mathtt c$, and the root above $\mathtt a$ and $u$, with $m(\mathrm{root})=u$.
For $w=\mathtt{acbabaac}$, we have $w_{|u}=\mathtt{cbbc}$, $B_{\mathrm{root}}(w)=01101001$ and $B_u(w)=1001$.
Conversely, $w$ is recovered by reading $B_{\mathrm{root}}$: a $0$ gives an $\mathtt a$, and a $1$ gives the next letter of $w_{|u}$, which is read off $B_u$.
\end{example}

As the example suggests, the node words determine the word, and they are independent and uniform when the word is uniform.

\begin{lemma}\label{lem:tree}
The map $w\mapsto(B_v(w))_{v\text{ internal}}$ is a bijection from $\W(k)$ onto $\prod_v\W(s_v-a_v,a_v)$.
Consequently, $\Ent=\sum_v\Ent_v$ with $\Ent_v=\lg\binom{s_v}{a_v}$, and if the node words are independent and each is uniform, the word that they encode is uniform on $\W(k)$.
\end{lemma}

\begin{proof}
Set $w_{|j}=j^{k_j}$ at a leaf $j$; at an internal node $v$ whose other child is $u$, the word $w_{|v}$ is obtained by reading $B_v(w)$ from left to right and taking the next letter of $w_{|m(v)}$ at a $1$ and the next letter of $w_{|u}$ at a $0$.
By induction from the leaves, this recovers $w$ from its node words, so that the map is injective.
The two sets have the same cardinality, since $\prod_v\binom{s_v}{a_v}=\prod_v\frac{s_v!}{a_v!\,(s_v-a_v)!}=\frac{n!}{\prod_jk_j!}$, each $s_v!$ of a non-root node $v$ appearing once in a numerator and once in a denominator.
\end{proof}

It therefore suffices to sample independent uniform binary words, one per internal node, each with a nearly entropic number of bits, and to assemble them.
Since the assembly costs a constant number of bit operations per letter of every node word, the total length $\sum_vs_v$ of the node words should be $O(n)$, which is why we use the Huffman tree~\cite{Huffman1952} of the counts.

\begin{lemma}\label{lem:huffman}
For the Huffman tree of $k_0,\dots,k_r$, $\sum_vs_v\le n(h+1)\le n(1+\lg(r+1))$, where $h=-\sum_j\frac{k_j}n\lg\frac{k_j}n$.
\end{lemma}

\begin{proof}
If $d_j$ is the depth of the leaf $j$, each occurrence of $j$ is counted at each of its $d_j$ internal ancestors, so that $\sum_vs_v=\sum_jk_jd_j$.
Huffman's tree minimises $\sum_jk_jd_j$ among prefix codes, and the Shannon lengths $\lceil-\lg(k_j/n)\rceil$ satisfy Kraft's inequality, which gives the bound.
\end{proof}

We now describe how a node word, a uniform word of $\W(s-a,a)$ with $1\le a\le s/2$, is sampled.
There are three cases, separated by the thresholds $T_n=\lceil n^{1/4}\rceil$ on the length $s$ and $s^{3/4}$ on the number $a$ of ones.

All the uniform integers are drawn from a \emph{recycling source} (\cref{sec:model}), which keeps the unused part of its previous draws as a uniform integer, and can take back uniform integers produced by the algorithm; each node has its own source.

Short words, with $s<T_n$, are \emph{small} nodes, and are sampled exactly: a uniform rank in $\{0,\dots,\binom sa-1\}$ is drawn from the source and decoded in the combinatorial number system.
This wastes $O(1)$ bits and costs $O(T_n^2)=O(\sqrt n)$ bit operations, which is affordable since there are at most $r$ such nodes.

Longer words with few ones, with $s\ge T_n$ and $a<s^{3/4}$, are \emph{sparse} nodes, and their ones are inserted one by one at uniform positions (\cref{alg:recycled} of \cref{sec:sparse}).
The Fisher--Yates shuffle would read about $\lg\binom sa+\lg a!$ bits, the extra $\lg a!$ bits being the ranks of the inserted ones among the ones.
The rank of each new one is uniform and independent of the new word, and we give it back to the source, where it pays for the next insertions; the insertions then cost exactly $\lg\binom sa$ bits, up to the small losses of the source.

Longer words with many ones, with $s\ge T_n$ and $a\ge s^{3/4}$, are \emph{dense} nodes, and are generated by \textsc{MergeShuffle} (\cref{alg:blockmerge} of \cref{sec:dense}): independent letters of law $(1-a/s,a/s)$ are drawn until one of the two letters exceeds its count, and the missing letters are then inserted at uniform positions, which is cheap since there are few of them when $a\ge s^{3/4}$.
The letters are drawn by blocks of $S$ letters carrying about $B=\lfloor\lg s/8\rfloor$ bits of entropy: the number of ones of the block is drawn by the Knuth--Yao sampler, and then its rank among the blocks with this number of ones is drawn from the source.
If $a/s<1/(B\lg B)$, a block has few ones, and the rank is decoded by locating only these ones in the combinatorial number system.
Otherwise, the number of ones of a block is concentrated around its mean, and the rank is decoded with a precomputed table of the blocks whose number of ones is typical, a table of size $O(\sqrt s\log^4s)$; the rare blocks of atypical type are decoded slowly (\cref{alg:block}).
\Cref{tab:samplers} summarises the guarantees of the three samplers, and \cref{alg:main} puts them together.

\begin{table}[htbp]
\centering
\begin{tabular}{@{}llll@{}}
\toprule
node & sampler & expected fresh bits & bit operations\\
\midrule
small, $s<T_n$ & exact & $\le\Ent_v+O(1)$ & $O(\sqrt n)$\\
sparse, $a<s^{3/4}$ & recycled insertions (\cref{thm:sparse}) & $\le\big(1+O(1/\log s)\big)\Ent_v$ & $O(s)$\\
dense, $a\ge s^{3/4}$ & blocks (\cref{thm:dense}) & $\le\big(1+O(1/\log s)\big)\Ent_v$ & $O(s)$\\
\bottomrule
\end{tabular}
\caption{The three samplers of node words, with $\Ent_v=\lg\binom sa$; the constants in the $O$'s are absolute, and the bounds for sparse and dense nodes hold for $s$ large enough.}\label{tab:samplers}
\end{table}

\begin{algorithm}[htbp]
\DontPrintSemicolon
\caption{\textsc{Words}$(k_0,\dots,k_r)$; each internal node uses its own recycling source}\label{alg:main}
build the Huffman tree of $(k_0,\dots,k_r)$, with deterministic ties; compute $s_v$, $m(v)$ and $a_v$ for its internal nodes\;
\For{each internal node $v$}{
  \uIf{$s_v<T_n$}{$B_v\leftarrow$ the word of $\W(s_v-a_v,a_v)$ whose ones form the $a_v$-subset of rank \textsc{Draw}$\big(\binom{s_v}{a_v}\big)$ in the combinatorial number system\;}
  \uElseIf{$a_v^4<s_v^3$}{$B_v\leftarrow$\textsc{RecycledInsertion}$(s_v-a_v,a_v)$ (\cref{alg:recycled})\;}
  \lElse{$B_v\leftarrow$\textsc{DenseMerge}$(s_v-a_v,a_v)$ (\cref{alg:blockmerge,alg:block})}
}
\For{each internal node $v$, in the order of its creation}{
  $w_{|v}\leftarrow$ the word read off $B_v$ from $w_{|m(v)}$ and $w_{|u}$, where $u$ is the other child of $v$ and $w_{|j}=j^{k_j}$ at a leaf $j$ (\cref{lem:tree})\;
}
\Return{$w_{|\mathrm{root}}$}\;
\end{algorithm}

Huffman's algorithm creates a node after its two children, so that the last loop of \cref{alg:main}, the \emph{assembly}, computes the inverse map of \cref{lem:tree}.
Our main theorem follows from the guarantees of \cref{tab:samplers}; it is proved in \cref{sec:proof}.

\begin{theorem}\label{thm:main}
There are absolute constants $n_0$ and $C$ such that, for every $n\ge n_0$, every $r\ge1$ and all positive integers $k_0,\dots,k_r$ with sum $n$, the following holds.
\begin{enumerate}[label=(\roman*)]
\item \Cref{alg:main} is a sampler for the uniform distribution on $\W(k_0,\dots,k_r)$.
\item The expected number $F$ of fresh bits it reads satisfies $\Ent\le\E F\le\big(1+C/\lg n\big)\Ent$.
\item It performs $O\big(n(1+\lg(r+1))^2+r\sqrt n\big)$ bit operations on average, with an absolute constant.
\end{enumerate}
\end{theorem}

The excess comes from three sources: the up to $2$ bits wasted by the Knuth--Yao sampler at each block of the dense nodes, the small loss of the recycling sources at each draw, and the $O(1)$ bits wasted by each small node.
For $k$-subsets, the tree has a single node, of size $n$, and we get the following.

\begin{corollary}\label{cor:subsets}
For $n$ large enough and every $1\le k\le n-1$, \textsc{Words}$(n-k,k)$ samples a uniform $k$-subset of $[n]$ with $O(n)$ bit operations on average, and reads on average at most $\big(1+O(1/\log n)\big)\lg\binom nk$ fresh bits, with absolute constants.
\end{corollary}

\section{Model and tools}\label{sec:model}

Before analysing the samplers of node words, we fix the model of computation and recall the classical tools on random bits that they use.
Logarithms are in base two unless written $\ln$, $[m]=\{1,\dots,m\}$, and $h_2(x)=-x\lg x-(1-x)\lg(1-x)$ is the binary entropy function.
For a vector $c=(c_0,\dots,c_s)$ of non-negative integers, $\W(c)$ is the set of words with exactly $c_j$ occurrences of the letter $j$, and $e_x$ is the vector with a single occurrence of $x$.
For integers $0\le k\le N$, $(N)_k=N!/(N-k)!$.

The memory is an unbounded array of bits, initially $0$, and an integer is stored as a string of consecutive bits together with its bit length, which is $\max(1,\lceil\lg(x+1)\rceil)$ for an integer $x\ge0$.
The machine has a constant number of registers, each holding an integer.
Reading or writing one bit of memory, at an address held in a register, costs $1$, whatever the length of this address.
Copying an integer between the memory and a register costs its bit length; in particular, an address stored in memory is loaded at the cost of its bit length, and computing an address is charged like any other arithmetic.
Adding, subtracting or comparing two integers of at most $L$ bits costs $O(L)$, and incrementing or decrementing an integer costs $1$ plus the number of its bits that change.
Multiplications and divisions with remainder of an integer of $L$ bits by an integer of $\ell\le L$ bits are performed by the schoolbook methods, at a cost of $O(L\ell)$.
Reading a fresh random bit costs $1$.
The cost of an algorithm is the total cost of its operations, its number of \emph{bit operations}.
This model differs from the random-access machine with logarithmic cost of Cook and Reckhow~\cite{CookReckhow1973} in that the length of an address is not charged.

An array of entries of $b$ bits is stored with each entry padded to $2^{\lceil\lg b\rceil}$ bits, at an address that is a multiple of a power of two at least equal to its size in bits; the address of its entry $i$ is obtained by writing $i$, followed by $\lceil\lg b\rceil$ zeros, in the low-order bits of the address of the array, held in a register, and these bits are cleared after the access.
A two-dimensional array is stored with rows of a power-of-two length, so that the index of an entry is the concatenation of its two indices.
A word over an alphabet of size $a\ge2$ is an array of letters of $\lceil\lg a\rceil$ bits.
The memory used by our algorithms is polynomial in $n$.
We use the following consequences of these conventions.

\begin{enumerate}[label=(B\arabic*)]
\item\label{B:counter} A counter of $\ell$ bits that is only incremented, or only decremented, $L$ times changes at most $2L+\ell$ bits in total: an increment changes $1+t$ bits, where $t$ is the number of trailing ones of the counter, and among $L$ consecutive integers at most $L/2^i+1$ have at least $i$ trailing ones; decrements are symmetric. In particular, a counter that is decremented until a decrement would make it negative, which is detected by the borrow leaving its top bit, costs $O(1)$ per decrement, amortised, plus $O(\ell)$.
\item\label{B:scan} By~\ref{B:counter} applied to the pointer, a sequential scan of $L$ consecutive entries of an array of entries of $b$ bits costs $O(Lb)$, plus the cost of the first access.
\item\label{B:access} An access to the entry $i$ of an array of entries of $b$ bits costs $O(\log(i+2)+b)$.
\end{enumerate}

In particular, an algorithm that writes the $n$ letters of its output performs $\Omega(n)$ bit operations.
The constants hidden in the bounds on bit operations depend on those of~\ref{B:counter}--\ref{B:access}, and are not tracked; all the constants on random bits are explicit.

We now turn to random bits.
An algorithm reads its random source sequentially, and decides whether to read a further bit from the bits already read; the number of bits that it reads is therefore a stopping time for the source.
When several procedures are run one after the other on the same source, each reads a segment delimited by stopping times, and, by the strong Markov property, the remainder of the source after a stopping time is a sequence of independent fair bits, independent of the bits already read.
A \emph{sampler} for a random variable $X$ is an algorithm that halts almost surely and whose output has the law of $X$.
This already gives the lower bound on the number of random bits.

\begin{lemma}[\cite{KnuthYao1976}]\label{lem:lower}
A sampler for $X$ that reads $F$ fresh bits satisfies $\E F\ge\Ent(X)$.
\end{lemma}

\begin{proof}
The words that the sampler may read form a prefix-free set, since the number of bits read is a stopping time, and a word $u$ is read with probability $2^{-|u|}$; the entropy of the word read is therefore $\E F$, and $X$ is a function of it.
\end{proof}

Distributions that are not uniform, such as the number of ones of a block, are drawn with the Knuth--Yao sampler, which reads the binary expansions of the probabilities.

\begin{lemma}[Knuth and Yao~\cite{KnuthYao1976}]\label{lem:ky}
Let $q$ be a probability distribution on a finite set $\mathcal X$, given through the binary expansions of the $q_x$.
The Knuth--Yao sampler reads at most $\Ent(q)+2$ bits on average, and more than $d$ bits with probability at most $|\mathcal X|2^{-d}$.
\end{lemma}

\begin{proof}
The first statement is the theorem of~\cite{KnuthYao1976}.
For the second, the sampler halts within $d$ bits with probability $\sum_x\lfloor2^dq_x\rfloor2^{-d}\ge1-|\mathcal X|2^{-d}$.
\end{proof}

We implement it by the following walk.
Fix an order on $\mathcal X$, and let $\mathcal L_d$ be the list of the $x$ whose $d$-th binary digit of $q_x$ is $1$.
Start with $\mathit{pos}=0$; for $d=1,2,\dots$, read a bit $b$, set $\mathit{pos}\leftarrow2\,\mathit{pos}+b$, return $\mathcal L_d[\mathit{pos}]$ if $\mathit{pos}<|\mathcal L_d|$, and otherwise set $\mathit{pos}\leftarrow\mathit{pos}-|\mathcal L_d|$.

Uniform integers, on the other hand, are drawn from a \emph{recycling source}, which keeps the unused part of the previous draws as a uniform integer instead of discarding it, and which can also take back uniform integers produced by the algorithm.
It extends the Fast Dice Roller of Lumbroso~\cite{Lumbroso2013}, which builds a uniform integer by doubling a smaller uniform integer with a fair bit: instead of starting each draw from scratch, the source starts from the uniform integer left by the previous draws.
A similar mechanism was developed independently by Draper and Saad~\cite{DraperSaad2026}, under the name of randomness recycling, in a word-RAM; we give a short self-contained analysis in the bit-RAM, where the refill threshold must be relative to the size of the draw.
We fix an integer $g\ge2$, say $g=8$.
The source holds a pair $(m,u)$ of integers with $0\le u<m$, initially $(1,0)$, and supports two operations.
\begin{itemize}
\item \textsc{Absorb}$(j,a)$, for an integer $0\le j<a$, replaces $(m,u)$ by $(am,au+j)$.
\item \textsc{Draw}$(L)$, for an integer $L\ge1$, repeats the following steps until it returns.
While $m<2^gL$, it reads a fresh bit $b$ and replaces $(m,u)$ by $(2m,2u+b)$.
It then writes $m=qL+r$ with $0\le r<L$.
If $u<qL$, it replaces $(m,u)$ by $(q,\lfloor u/L\rfloor)$ and returns $u\bmod L$, computed from the former value of $u$; otherwise, it replaces $(m,u)$ by $(r,u-qL)$.
\end{itemize}
We call \emph{visible} all the data of a computation that uses the source, except $u$.

\begin{lemma}\label{lem:source}
Consider a computation that uses a recycling source, and assume that each absorbed integer $j$, of size $a$, is uniform on $\{0,\dots,a-1\}$ and independent of $u$, conditionally on the visible data just after the absorption.
\begin{enumerate}[label=(\roman*)]
\item At all times, conditionally on the visible data, $u$ is uniform on $\{0,\dots,m-1\}$; in particular, each \textsc{Draw}$(L)$ returns, conditionally on the visible data before it, a uniform integer of $\{0,\dots,L-1\}$.
\item The expected number of fresh bits read by the source is at most
\[
\E\Big[\sum\lg L-\sum\lg a+\lg m_{\mathrm{end}}\Big]+2(g+1)2^{-g}\,\E[\text{number of draws}],
\]
where the sums are over the draws and the absorptions, and $m_{\mathrm{end}}$ is the final value of $m$.
\item Assume that every absorption of size $a$ immediately follows a draw of size at least $a$, and that every draw has a size at least $A$, the largest absorbed size so far ($A=1$ if there is none).
Then $m<2^{g+1}L$ during a draw of size $L$, and $m<2^{g+1}A$ between two operations.
A draw of size $L$ costs on average $O\big(g\log(L+1)\big)$ bit operations, plus $O(1)$ per fresh bit read, and an absorption of size $a$ costs $O\big((g+\log A)\log(a+1)\big)$.
\end{enumerate}
\end{lemma}

\begin{proof}
For (i), we check that each step preserves the property.
Reading a fresh bit $b$ gives $2u+b$, uniform on $\{0,\dots,2m-1\}$.
At a test, conditionally on $u<qL$, the integer $u$ is uniform on $\{0,\dots,qL-1\}$, so that $u\bmod L$ and $\lfloor u/L\rfloor$ are independent and uniform on $\{0,\dots,L-1\}$ and $\{0,\dots,q-1\}$; conditionally on $u\ge qL$, $u-qL$ is uniform on $\{0,\dots,r-1\}$.
An absorption gives $au+j$, uniform on $\{0,\dots,am-1\}$ by the assumption.
Finally, the other steps of the computation only modify the visible data, using the visible data and fresh bits independent of $u$, or forget part of them, which preserves the property.

For (ii), consider the potential $\lg m$.
A fresh bit increases it by $1$, and an absorption by $\lg a$.
At a test, let $P=qL/m$ be the probability that $u<qL$; since $u$ is uniform on $\{0,\dots,m-1\}$, which is split into $q$ blocks of size $L$ and a remainder of size $r$, we have $\lg m=h_2(P)+P\lg(qL)+(1-P)\lg r$, so that the potential decreases on average by $P\lg L+h_2(P)$.
Over a draw, the terms $P\lg L$ add up to $\lg L$, since exactly one test succeeds.
At each test, $m\ge2^gL$ and $r<L$, so that $1-P<2^{-g}\le1/4$; hence $h_2(P)=h_2(1-P)\le2^{-g}(g+1.45)$, using $-(1-x)\lg(1-x)\le x\lg e$, and each test fails with probability less than $1/4$, so that a draw makes on average at most $4/3$ tests.
Summing the variations of the potential, which starts at $0$, gives (ii).

For (iii), during a draw, fresh bits are read only while $m<2^gL$, so that $m<2^{g+1}\max(A,L)=2^{g+1}L$.
After a successful test, $q\le m/L<2^{g+1}$, and a following absorption of size $a\le L$ gives $m<2^{g+1}a\le2^{g+1}A$; after a failed test, $m=r<L$, and the draw goes on.
The integers handled during a draw of size $L$ therefore have $O(g+\log(L+1))$ bits, and the quotients computed at a test have $O(g)$ bits, so that a test costs $O(g\log(L+1))$ by long division; a fresh bit is appended to $m$ and $u$ at a cost of $O(1)$, amortised by~\ref{B:counter}.
An absorption multiplies integers of $O(g+\log A)$ and $O(\log(a+1))$ bits.
\end{proof}

Both samplers of binary words also insert letters at uniform positions.
Given a word $W$ of length $L$ and a letter $x$, \Insert{$W,x$} draws a position $p$ by \textsc{Draw}$(L+1)$, appends $x$ to $W$, and exchanges the letters at positions $p$ and $L$, positions being counted from $0$.
Besides the new word $W'$, it defines the rank $j$ of $p$ among the positions of $x$ in $W'$, that is, the number of occurrences of $x$ in $W'$ before position $p$.
The following lemma says that this rank, which the Fisher--Yates shuffle generates for nothing, is uniform and independent of the new word, so that it can be given back to the source.

\begin{lemma}\label{lem:insert}
If $W$ is uniform on $\W(c)$, of length $L$, and $p$ is uniform on $\{0,\dots,L\}$ and independent of $W$, then $W'$ is uniform on $\W(c+e_x)$, and $j$ is uniform on $\{0,\dots,c_x\}$ and independent of $W'$.
\end{lemma}

\begin{proof}
The map $(W,p)\mapsto(W',j)$ is a bijection from $\W(c)\times\{0,\dots,L\}$ onto $\W(c+e_x)\times\{0,\dots,c_x\}$: given $W'$ and $j$, the position $p$ is that of the $(j+1)$-th occurrence of $x$ in $W'$, and $W$ is obtained by exchanging the positions $p$ and $L$ of $W'$ and deleting the last letter.
Indeed, both sets have $|\W(c)|(L+1)=|\W(c+e_x)|(c_x+1)$ elements.
\end{proof}

Finally, we shall compare the number of bits read with $\Ent$ through the following classical estimates: for $0\le k\le N$, $\lg\binom Nk\ge k\lg(N/k)$, and, by the method of types~\cite[Thm.~11.1.3]{CoverThomas2006},
\begin{equation}\label{eq:types}
Nh_2(k/N)-2\lg(N+1)\le\lg\binom Nk\le Nh_2(k/N).
\end{equation}

\section{Sparse nodes: insertions with recycling}\label{sec:sparse}

In this section, $n$ and $k$ are integers with $\lambda=\lg n\ge16$ and $1\le k<n^{3/4}$; the target set is $\W(n-k,k)$, with $\Ent=\lg\binom nk$.
In \cref{alg:main}, $n$ and $k$ are the length $s_v$ and the number of ones $a_v$ of a sparse node.

Inserting the $k$ ones one by one at uniform positions into $0^{n-k}$, as the Fisher--Yates shuffle does, reads about $\lg(n)_k=\Ent+\lg k!$ bits.
The extra $\lg k!$ bits are the ranks of the inserted ones: when the $(t+1)$-th one is inserted, its rank $j$ among the ones is uniform on $\{0,\dots,t\}$ and independent of the new word (\cref{lem:insert}).
We give this rank back to the source, which uses it for the next insertions.
The net cost of the $(t+1)$-th insertion is then $\lg(L+1)-\lg(t+1)$ bits, where $L=n-k+t$ is the current length, and these costs add up exactly to $\Ent$.

\begin{algorithm}[htbp]
\DontPrintSemicolon
\caption{\textsc{RecycledInsertion}$(n-k,k)$}\label{alg:recycled}
initialise a recycling source; $W\leftarrow0^{n-k}$\;
\For{$t\leftarrow0$ \KwTo $k-1$}{
  $(W,j)\leftarrow$\Insert{$W,1$} \tcp*[f]{$j$ is the rank of the new one}\;
  \textsc{Absorb}$(j,t+1)$\;
}
\Return{$W$}\;
\end{algorithm}

\begin{theorem}\label{thm:sparse}
Let $\lambda=\lg n\ge16$ and $1\le k<n^{3/4}$.
\Cref{alg:recycled} is a sampler for the uniform distribution on $\W(n-k,k)$ that reads on average at most
\[
\lg\binom nk+2(g+1)2^{-g}k+g+1+\lg k=\big(1+O(1/\log n)\big)\lg\binom nk
\]
fresh bits, and performs $O(n)$ bit operations on average.
\end{theorem}

\begin{proof}
We first show by induction on $t$ that, after $t$ insertions, $W$ is uniform on $\W(n-k,t)$, conditionally on the visible data other than $W$.
This holds for $t=0$.
At the next insertion, the position $p$ is uniform and independent of $W$ by \cref{lem:source}(i), so that, by \cref{lem:insert}, the new word $W'$ is uniform and the rank $j$ is uniform on $\{0,\dots,t\}$ and independent of $W'$.
Once $W'$ and $j$ are computed, the algorithm forgets $W$ and $p$; since $u$ was independent of $(W,p)$, it is independent of $(W',j)$, and the assumption of \cref{lem:source} holds for the absorption of $j$.
This proves that the output is uniform.

By \cref{lem:source}(ii), the expected number of fresh bits is at most
\[
\sum_{t=0}^{k-1}\lg(n-k+t+1)-\sum_{t=0}^{k-1}\lg(t+1)+\E\lg m_{\mathrm{end}}+2(g+1)2^{-g}k,
\]
and the first two sums give $\lg(n)_k-\lg k!=\Ent$.
The assumptions of \cref{lem:source}(iii) hold, since the absorbed sizes are at most $k\le n-k+1$, so that $m_{\mathrm{end}}<2^{g+1}k$.
This gives the bound, and the right-hand side is $\Ent+O(k)$, while $\Ent\ge k\lg(n/k)>k\lambda/4$.

For the bit operations, the word $W$ is an array of $n$ bits, and a complete binary tree of counters over the $n$ positions, stored in heap order, records the number of ones below each node; since the memory is initially zero, neither needs initialisation.
An insertion updates the counters along one or two paths, and $j$ is the sum of the counters of the left siblings along the path of $p$; each path has $O(\log n)$ nodes, which cost $O(\log n)$ each by~\ref{B:access}.
By \cref{lem:source}(iii), the draws and the absorptions cost $O(\log^2n)$ each on average, including the fresh bits read, so that the $k<n^{3/4}$ insertions cost $O(k\log^2n)=O(n)$.
\end{proof}

\section{Dense nodes: \textsc{MergeShuffle} driven by blocks}\label{sec:dense}

In this section, $c_0\ge c_1\ge N^{3/4}$ are integers with $c_0+c_1=N$ and $\lg N\ge512$; we write $p=c_1/N\in[N^{-1/4},1/2]$, $h=h_2(p)$ and $\Ent_N=\lg\binom N{c_1}$.
In \cref{alg:main}, $N=s_v$ and $c_1=a_v$ for a dense node.

\textsc{MergeShuffle}, in the form of \cref{alg:blockmerge}, draws independent letters of law $(1-p,p)$ until one of the two letters exceeds its count, and inserts the missing letters at uniform positions.
Whatever the way in which the letters are drawn, the output is uniform, and the number of insertions is small when $c_1\ge N^{3/4}$.

\begin{algorithm}[htbp]
\DontPrintSemicolon
\caption{\textsc{BlockMerge}$(c_0,c_1)$}\label{alg:blockmerge}
$w\leftarrow$ empty word; $(R_0,R_1)\leftarrow(c_0,c_1)$; $\mathit{buf}\leftarrow$ empty word\;
\While{\textnormal{\textbf{true}}}{
  \lIf{$\mathit{buf}$ is empty}{$\mathit{buf}\leftarrow$ a block drawn by the block sampler}
  $x\leftarrow$ first letter of $\mathit{buf}$, removed from $\mathit{buf}$\;
  \lIf{$R_x=0$}{\textbf{break}}
  $w\leftarrow w\cdot x$; $R_x\leftarrow R_x-1$\;
}
\lFor{$x\in\{0,1\}$ and $t\leftarrow 1$ \KwTo $R_x$}{\Insert{$w,x$}}
\Return{$w$}\;
\end{algorithm}

\begin{lemma}\label{lem:blockmerge}
If the blocks consist of independent letters of law $(1-p,p)$, and the positions of the insertions are drawn from a recycling source, \cref{alg:blockmerge} is a sampler for the uniform distribution on $\W(c_0,c_1)$, and its expected number $M$ of final insertions satisfies $\E M\le2N(1/c_0+1/c_1)^{1/2}+1\le2\sqrt2N^{5/8}+1$.
\end{lemma}

\begin{proof}
The letters examined by the loop are the first letters of a sequence $X_1,X_2,\dots$ of independent letters of law $(1-p,p)$, and the loop stops at the first letter $X_T$ that would exceed its quota, so that $T\le N+1$.
For a word $u$ with $a_0\le c_0$ zeros and $a_1\le c_1$ ones and a letter $x$ with $a_x=c_x$, the event that the loop stops with $w=u$ and overflowing letter $x$ is $\{X_1\cdots X_{|u|+1}=u\cdot x\}$, of probability $(1-p)^{a_0}p^{a_1}\Prob(X_1=x)$, which does not depend on the order of the letters of $u$; conditionally on $(a_0,a_1,x)$, the word $w$ is uniform.
The positions of the final insertions are uniform conditionally on everything computed before them (\cref{lem:source}(i)), so that, by \cref{lem:insert}, the output is uniform conditionally on $(a_0,a_1,x)$, hence unconditionally.

For the insertions, note that $M=N+1-T\le N$, so that $\Prob(M\ge u)=0$ for $u>N$.
Extend the sequence $(X_t)$ beyond $T$, and let $1\le u\le N$ and $t=N+1-u$.
The event $\{M\ge u\}=\{T\le t\}$ means that some letter $j$ occurs at least $c_j+1$ times among $X_1,\dots,X_t$.
This number of occurrences is binomial with mean $c_j-(u-1)c_j/N$ and variance at most $c_j$, so that it exceeds its mean by at least $uc_j/N$, and Chebyshev's inequality gives $\Prob(M\ge u)\le\frac{N^2}{u^2}\sum_j\frac1{c_j}$.
With $A=N(1/c_0+1/c_1)^{1/2}$, summing $\min(1,A^2/u^2)$ over $u\ge1$ gives at most $2A+1$.
Finally $1/c_0+1/c_1\le2N^{-3/4}$.
\end{proof}

The letters themselves cannot be drawn one by one: when $p$ is small, a letter carries $h\ll1$ bits of entropy, and drawing it costs at least one bit.
We therefore draw them by blocks of $S$ letters, carrying about $B$ bits of entropy each, where
\[
B=\Big\lfloor\frac{\lg N}8\Big\rfloor\ge64,\qquad S=\lceil B/h\rceil,\qquad\delta=\frac1{B\lg B}\le\frac1{384}.
\]
A block consists of $S$ letters, and carries $Sh\in[B,B+1)$ bits of entropy.
It is drawn in two steps.
First, its \emph{type}, that is, its number $A$ of ones, of law $\Prob(A=t)=\binom St p^t(1-p)^{S-t}$, is drawn by the Knuth--Yao walk.
Then its rank among the blocks of this type is drawn from the recycling source of the node, and decoded; we write $\vartheta_A=\lg\binom SA$.
For $0\le t\le S$ and $0\le\rho<\binom St$, let $\mathrm{word}(t,\rho)$ be the binary word of length $S$ whose ones form the $t$-subset $\{u_1<\dots<u_t\}$ of $\{0,\dots,S-1\}$ of rank $\rho=\sum_{i=1}^t\binom{u_i}i$ in the combinatorial number system~\cite[Sec.~7.2.1.3]{KnuthTAOCP4A} if $t\le S/2$, and the complement of $\mathrm{word}(S-t,\rho)$ otherwise; the map $\rho\mapsto\mathrm{word}(t,\rho)$ is a bijection from $\{0,\dots,\binom St-1\}$ onto $\W(S-t,t)$.
The decoding of the rank is the delicate part, since it handles integers of about $B$ bits.
If $p<\delta$, a block has few ones, and only their positions are computed, by binary searches in the combinatorial number system; we call this the \emph{sparse decoding}.
If $p\ge\delta$, the blocks are short, and their number of ones is concentrated around its mean $\mu=pS$.
We call a type $t$ \emph{typical} if $\binom St\le2^{4B+4}$, and we write $\mathcal T$ for the set of typical types; $\mathrm{word}(A,\rho)$ is then read in a precomputed table of the words $\mathrm{word}(t,\rho')$ with $t\in\mathcal T$, and the rare blocks of atypical type are decoded by the sparse decoding.
\textsc{DenseMerge}$(c_0,c_1)$ is \cref{alg:blockmerge} with the block sampler of \cref{alg:block}.

\begin{algorithm}[htbp]
\DontPrintSemicolon
\caption{The block sampler for $(c_0,c_1)$}\label{alg:block}
$A\leftarrow$ a type, drawn by the Knuth--Yao walk on $\big(\binom St p^t(1-p)^{S-t}\big)_{0\le t\le S}$\;
$\rho\leftarrow$\textsc{Draw}$\big(\binom SA\big)$\;
\eIf(\tcp*[f]{table decoding}){$p\ge\delta$ and $A\in\mathcal T$}{
  \Return{the entry $(A,\rho)$ of the table of the words $\mathrm{word}(t,\rho')$, $t\in\mathcal T$}\;
}(\tcp*[f]{sparse decoding}){
  $a\leftarrow\min(A,S-A)$; $\mathit{hi}\leftarrow S$\;
  \For{$i\leftarrow a$ \KwTo $1$}{
    $u_i\leftarrow$ the largest $u\in\{i-1,\dots,\mathit{hi}-1\}$ with $\binom ui\le\rho$, by binary search\;
    $\rho\leftarrow\rho-\binom{u_i}i$; $\mathit{hi}\leftarrow u_i$\;
  }
  \Return{the word with the letter $[A\le S/2]$ at the positions $u_1,\dots,u_a$ and the other letter elsewhere}\;
}
\end{algorithm}

The greedy steps of the sparse decoding compute the subset of rank $\rho$ in the combinatorial number system; since $\binom{i-1}i=0\le\rho$, each binary search is well defined.
The block sampler draws exactly the right letters.

\begin{lemma}\label{lem:block}
Conditionally on the previous blocks, a block consists of $S$ independent letters of law $(1-p,p)$.
Its type is drawn with at most $\Ent(A)+2$ fresh bits on average, and $\Ent(A)+\E\vartheta_A=Sh$.
\end{lemma}

\begin{proof}
Under the law of $S$ independent letters of law $(1-p,p)$, the number of ones has the law of $A$ and, conditionally on it, the word is uniform on $\W(S-A,A)$.
The type is drawn with fresh bits, independent of the previous blocks, and, by \cref{lem:source}(i), $\rho$ is uniform conditionally on the previous blocks and on $A$; since $\rho\mapsto\mathrm{word}(A,\rho)$ is a bijection, the block has the required law.
The bound on the fresh bits of the type is \cref{lem:ky}, and $\Ent(A)+\E\vartheta_A=Sh$ is the entropy of the block, which is uniform given its type.
\end{proof}

We now bound the cost of a block.
The first lemma bounds the length of the blocks, and the second one shows that, when the table is used, it is small and atypical blocks are rare.

\begin{lemma}\label{lem:sizes}
$S\le4BN^{1/4}/\lg N+1\le N^{1/4}$, and, if $p\ge\delta$, $S\le B^2+1$.
\end{lemma}

\begin{proof}
The function $h_2$ is increasing on $(0,1/2]$ and $h_2(x)\ge x\lg(1/x)$, so that $h\ge h_2(N^{-1/4})\ge\frac14N^{-1/4}\lg N$; with $B\le\lg N/8$, this gives the first bound.
If $p\ge\delta$, then $h\ge\delta\lg(1/\delta)=\lg(B\lg B)/(B\lg B)\ge1/B$, so that $S\le B^2+1$.
\end{proof}

\begin{lemma}\label{lem:typical}
We have $\sum_{t\in\mathcal T}\binom St\le(S+1)2^{4B+4}$, and $2^{4B}\le\sqrt N$.
If $p\ge\delta$, then $\Prob(A\notin\mathcal T)\le B^{-3}$.
\end{lemma}

\begin{proof}
The first bound holds because each typical type has at most $2^{4B+4}$ words, and the second because $B\le\lg N/8$.
Assume $p\ge\delta$; we first show that an atypical type exceeds four times the mean $\mu=pS$.
Recall that $\binom St\le2^{Sh_2(t/S)}$ by~\eqref{eq:types}, and that $Sh<B+1$.
If $p>1/8$, then $h>h_2(1/8)>1/2$, so that $\binom St\le2^S<2^{2Sh}<2^{4B+4}$ for every $t$, and every type is typical.
If $p\le1/8$ and $t\le4\mu=4pS\le S/2$, then, since $h_2$ is increasing on $[0,1/2]$, and concave with $h_2(0)=0$, so that $h_2(4p)\le4h_2(p)$, we get $\binom St\le2^{Sh_2(4p)}\le2^{4Sh}<2^{4B+4}$, and $t$ is typical.
Hence $\Prob(A\notin\mathcal T)\le\Prob(A\ge4\mu)$.
By Chernoff's bound $\Prob(A\ge(1+\eta)\mu)\le\big(e^\eta/(1+\eta)^{1+\eta}\big)^\mu$ with $\eta=3$, we get $\Prob(A\ge4\mu)\le(e^3/4^4)^\mu\le e^{-2.5\mu}$.
It remains to bound $\mu$ from below.
Since $-(1-p)\lg(1-p)\le p\lg e$, we have $h\le p\lg(e/p)$, so that $\mu=pS\ge pB/h\ge B/\lg(e/p)\ge B/\lg(e/\delta)=B/\lg(eB\lg B)$.
For $B\ge64$, we have $e\lg B\le B$, hence $\lg(eB\lg B)\le2\lg B$ and $\mu\ge B/(2\lg B)$.
Hence $\Prob(A\notin\mathcal T)\le e^{-1.25B/\lg B}$, which is at most $B^{-3}$ for $B\ge64$.
\end{proof}

\begin{lemma}\label{lem:decode}
The tables described below cost $o(N)$ bit operations, and, given them, a block costs $O(S)$ bit operations on average, with absolute constants.
\end{lemma}

\begin{proof}
We start with the tables, which are computed once.
The type probabilities are $N_t/N^S$ with $N_t=\binom St c_1^tc_0^{S-t}$, integers of $O(S\log N)$ bits, computed from one another by multiplications and exact divisions by integers of $O(\log N)$ bits; their first $D=\lceil2\lg N\rceil$ binary digits are produced by long division, and give the lists $\mathcal L_1,\dots,\mathcal L_D$ of the Knuth--Yao walk.
The binomial coefficients $\binom ui$, $0\le i\le u\le S$, are computed by Pascal's rule.
By \cref{lem:sizes}, all this costs $O(S^3+S^2\log^2N)=O(N^{3/4})$.
If $p\ge\delta$, the table of the words $\mathrm{word}(t,\rho)$, $t\in\mathcal T$, is stored in an array indexed by the concatenation of $t$ and of $\rho$ written on $4B+4$ bits, together with a bit vector recording the typical types; each entry is computed by unranking, with $O(S^2)$ bit operations, so that, by \cref{lem:sizes,lem:typical}, the table costs $O(B^6\sqrt N)=o(N)$.
Deeper digits of the type probabilities are computed only when a walk needs them, at a cost of $O(S^2\log N)$ per level; since a walk reads more than $d$ bits with probability at most $(S+1)2^{-d}$ (\cref{lem:ky}), and at most $N+2$ walks are performed (\cref{lem:blockcount} below), this costs $O(S^3\log N/N)=o(1)$ on average.

We now turn to the cost of one block.
In the Knuth--Yao walk, the level $d$ is kept in a counter that is only incremented, and each bit read costs $O(\log(S+2))$ amortised by~\ref{B:counter} and~\ref{B:access}, and $O(\log(S+2))$ bits are read on average (\cref{lem:ky}), so that the walk costs $O(S)$ on average.

By \cref{lem:source}(iii), drawing the rank $\rho$ costs $O(g(\vartheta_A+1))=O(S)$ on average, including the fresh bits that it reads, since the source of a dense node absorbs nothing and $\vartheta_A\le S$.

It remains to decode the rank.
When the table is used, writing the index of the entry and copying its $S$ bits cost $O(S)$, since $S\ge B$.

We now bound the cost of the sparse decoding, for a block of type $A$; let $a=\min(A,S-A)$.
All the binomial coefficients compared with $\rho$ are of the form $\binom ui$ with $u\le S-1$ and $i\le a\le S/2$, so that they are at most $\binom Si\le\binom Sa=\binom SA$, and $\rho<\binom SA$: these integers have at most $\vartheta_A+1$ bits.
Each binary search makes $O(\log(S+2))$ comparisons, each of which accesses the table of binomial coefficients at a cost of $O(\log(S+2))$ for the index and compares integers of at most $\vartheta_A+1$ bits; each step also performs one subtraction.
Writing the block, a sequential write of $S$ bits followed by the $a$ positions $u_i$, costs $O(S+a\log(S+2))$.
Since $\vartheta_A\le a\lg S$, $\vartheta_A\le S$ and $a\le\min(A,S)$, the sparse decoding costs
\[
O\big(S+a\log(S+2)(\vartheta_A+\log(S+2))\big)=O\Big(S+\min\big(A(A+1),S^2\big)\log^2(S+2)\Big).
\]
If $p\ge\delta$, it is only used for atypical types, with probability at most $B^{-3}$ by \cref{lem:typical}, and costs then $O(S^2\log^2(S+2))=O\big(S(B^2+1)\log^2(B^2+3)\big)$ by \cref{lem:sizes}, hence $O(S)$ on average.
If $p<\delta$, it is used for every block; with $\mu=\E A=pS$, we have $\E[A(A+1)]\le\mu^2+2\mu$, so that the decoding costs on average $O(S)$ times $1+p(\mu+2)\lg^2(S+2)$.
Since $h\ge p\lg(1/p)$, we have $S\le B/(p\lg(1/p))+1$, hence $\mu\le B/\lg(1/p)+1$; and $S+2\le B/p+3\le2B/p$, so that $\lg(S+2)\le1+\lg B+\lg(1/p)\le2\lg(1/p)$, because $1/p>B\lg B\ge2B$.
Therefore
\[
p(\mu+2)\lg^2(S+2)\le4p\big(B\lg(1/p)+3\lg^2(1/p)\big).
\]
The functions $x\lg(1/x)$ and $x\lg^2(1/x)$ are increasing on $(0,e^{-2}]$, and $p<\delta<e^{-2}$, so that this is at most
\[
4\delta\big(B\lg(1/\delta)+3\lg^2(1/\delta)\big)=4\Big(\frac{\lg(B\lg B)}{\lg B}+\frac{3\lg^2(B\lg B)}{B\lg B}\Big)=O(1).
\]

In all cases, adding these bounds, a block costs $O(S)$ bit operations on average.
\end{proof}

Since the loop of \textsc{MergeShuffle} stops at a random time, the cost of the blocks is obtained from the cost of one block by the following argument of Wald's type.

\begin{lemma}\label{lem:blockcount}
The loop of \textsc{DenseMerge} draws at most $(N+1)/S+1$ blocks, and the event that a block is drawn depends only on the previous blocks.
Consequently, for every quantity $X_i$ attached to the $i$-th block whose conditional expectation given the previous blocks is at most $x$, the expected sum of the $X_i$ over the blocks drawn is at most $\big((N+1)/S+1\big)x$.
\end{lemma}

\begin{proof}
The loop examines at most $N+1$ letters, and it draws a new block only when the previous ones are exhausted.
Since the decision to draw the $i$-th block is a function of the first $i-1$ blocks, the expected sum is at most $\sum_i\Prob(\text{the $i$-th block is drawn})\,x$.
\end{proof}

Putting these lemmas together gives the guarantees of the sampler of dense nodes.

\begin{theorem}\label{thm:dense}
Let $c_0\ge c_1\ge N^{3/4}$ be integers with $c_0+c_1=N$ and $\lg N\ge512$, and $B=\lfloor\lg N/8\rfloor$.
Then \textsc{DenseMerge}$(c_0,c_1)$ is a sampler for the uniform distribution on $\W(c_0,c_1)$ that reads on average at most $\big(1+O(1/\log N)\big)\lg\binom N{c_1}$ fresh bits, and performs $O(N)$ bit operations on average, with absolute constants.
\end{theorem}

\begin{proof}
The output is uniform by \cref{lem:block,lem:blockmerge}.
As for the bits, the types read fresh bits directly, and the ranks and the positions of the insertions are drawn from the source of the node, which absorbs nothing, so that $m_{\mathrm{end}}<2^{g+1}$ by \cref{lem:source}(iii).
Let $c_g=2(g+1)2^{-g}$.
By \cref{lem:source}(ii), the source reads on average at most $\E\big[\sum_i(\vartheta_{A_i}+c_g)\big]+\E[M](\lg(N+1)+c_g)+g+1$ fresh bits, where $A_i$ is the type of the $i$-th block, and the types read at most $\Ent(A)+2$ fresh bits per block on average (\cref{lem:block}).
Since $\E[\vartheta_{A_i}]=\E\vartheta_A$ conditionally on the previous blocks, and $\Ent(A)+\E\vartheta_A=Sh$, \cref{lem:blockcount} shows that the blocks cost at most $\big((N+1)/S+1\big)(Sh+2+c_g)$ fresh bits on average.
Since $B/h\le S<B/h+1$ and $h\le1$, we have $1/S\le h/B$ and $Sh<B+1$, so that this is at most $\big(1+(2+c_g)/B\big)(N+1)h+B+3+c_g$.
By~\eqref{eq:types}, $(N+1)h\le Nh+1\le\Ent_N+2\lg(N+1)+1$.
By \cref{lem:blockmerge}, $\E[M](\lg(N+1)+c_g)=O(N^{5/8}\log N)$.
Altogether, the expected number of fresh bits is at most $\big(1+(2+c_g)/B\big)\Ent_N+O(N^{5/8}\log N)$.
Since $c_1\le N/2$, $\Ent_N\ge c_1\lg(N/c_1)\ge c_1\ge N^{3/4}$, so that the last term is $O(N^{-1/8}\log N)\,\Ent_N$; with $(2+c_g)/B=O(1/\log N)$, this gives the bound.

We turn to the bit operations.
The tables cost $o(N)$ (\cref{lem:decode}).
By \cref{lem:decode,lem:blockcount}, the blocks cost $O\big(S((N+1)/S+1)\big)=O(N)$ on average, since $S\le N$ (\cref{lem:sizes}).
In the loop, the quotas $R_0,R_1$ are counters that are only decremented, the test $R_x=0$ being the detection of~\ref{B:counter}, and the letters are read from the buffer and appended to $w$ sequentially; by~\ref{B:counter} and~\ref{B:scan}, the loop costs $O(N)$ apart from the blocks.
Each final insertion costs $O(\log N)$ on average (\cref{lem:source}(iii)), and there are $O(N^{5/8})$ of them on average (\cref{lem:blockmerge}).
\end{proof}

\section{Proof of the main theorem}\label{sec:proof}

It remains to put the pieces together.
The only new ingredient is a lower bound on $\Ent$, which shows that the $O(1)$ bits wasted by each small node are negligible.

\begin{lemma}\label{lem:entropy_r}
$\Ent\ge\lg(n)_r$, and $r\le2\Ent/\lg(n/2)$ if $n\ge4$.
\end{lemma}

\begin{proof}
If two counts satisfy $k\ge l\ge2$, replacing them by $k+1$ and $l-1$ multiplies $\prod_jk_j!$ by $(k+1)/l>1$; hence, for positive counts, $\prod_jk_j!\le(n-r)!$, and $\Ent=\lg n!-\sum_j\lg k_j!\ge\lg(n)_r$.
If $n-r+1\ge n/2$, this gives $\Ent\ge r\lg(n/2)$.
Otherwise, all the integers of $[n/2,n]$, of which there are at least $n/2$, are factors of $(n)_r$, so that $\Ent\ge\frac n2\lg\frac n2\ge\frac r2\lg\frac n2$.
In both cases $r\le2\Ent/\lg(n/2)$.
\end{proof}

\begin{proof}[Proof of \cref{thm:main}]
We take $n_0=2^{2048}$, so that every node with $s_v\ge T_n$ has $\lg s_v\ge\frac14\lg n\ge512$, and \cref{thm:sparse,thm:dense} apply.

We first check that the output is uniform.
Each node has its own recycling source, and the node samplers read disjoint segments of the random source delimited by stopping times; each node word is a function of the bits read by its sampler, so that the node words are independent.
They are uniform: for a small node by \cref{lem:source}(i) and the bijectivity of the combinatorial number system, for a sparse node by \cref{thm:sparse}, and for a dense node by \cref{thm:dense}.
By \cref{lem:tree}, the word that they encode is uniform, and the assembly computes it.

We now bound the number of fresh bits, node by node.
A small node makes a single draw, of size $\binom{s_v}{a_v}$, and absorbs nothing, so that it reads at most $\Ent_v+g+1+2(g+1)2^{-g}=\Ent_v+O(1)$ fresh bits on average (\cref{lem:source}).
A sparse or dense node reads at most $\big(1+O(1/\log s_v)\big)\Ent_v=\big(1+O(1/\log n)\big)\Ent_v$ fresh bits (\cref{thm:sparse,thm:dense}), since $s_v\ge n^{1/4}$.
There are at most $r$ small nodes, and $r\le2\Ent/\lg(n/2)$ by \cref{lem:entropy_r}.
Since $\Ent=\sum_v\Ent_v$ (\cref{lem:tree}), summing gives (ii); the lower bound is \cref{lem:lower}.

Finally, we bound the number of bit operations.
The counts are sorted by counting: a bit vector of length $n$, initially zero, marks the values $k_j$, the letters of each value are chained in a list, at a cost of $O(\log n)$ per letter, and a scan of the bit vector, of cost $O(n)$, lists the letters by increasing count.
Huffman's tree is then built by the method of the two queues, the second one receiving the created nodes in non-decreasing order of weight, with $O(1)$ comparisons and one addition of integers of $O(\log n)$ bits per node; the classification compares $a_v^4$ with $s_v^3$, at a cost of $O(\log^2s_v)=O(s_v)$.
This costs $O(n+r\log n+\sum_vs_v)$, and $r\log n=O\big(n(1+\lg(r+1))\big)$: if $r+1\ge\sqrt n$, then $\lg n\le2\lg(r+1)$ and $r\le n$; otherwise $r\lg n\le\sqrt n\lg n\le2n$.
For the small nodes, the binomial coefficients $\binom ui$, $0\le i\le u<T_n$, are computed once by Pascal's rule, at a cost of $O(T_n^3)=O(n^{3/4})$; at a small node, the draw costs $O(T_n)$ on average (\cref{lem:source}(iii)), and the unranking performs $O(s_v)$ comparisons and subtractions of integers of $O(T_n)$ bits, hence $O(T_n^2)=O(\sqrt n)$ per node.
A sparse or dense node costs $O(s_v)$ on average (\cref{thm:sparse,thm:dense}), and each node costs $O(\log n)$ more to set its parameters and the address of its word.
At the assembly, the node $v$ reads $B_v$, $w_{|m(v)}$ and $w_{|u}$ sequentially and writes $w_{|v}$, with $\lceil\lg(r+1)\rceil$ bits per letter, at a cost of $O(s_v\lceil\lg(r+1)\rceil+\log n)$ by~\ref{B:counter} and~\ref{B:scan}.
By \cref{lem:huffman}, $\sum_vs_v\le n(1+\lg(r+1))$, and the total is $O\big(n(1+\lg(r+1))^2+r\sqrt n\big)$.
\end{proof}

\begin{proof}[Proof of \cref{cor:subsets}]
The tree has a single internal node, of size $n\ge T_n$, so that there is no small node.
If $\min(k,n-k)<n^{3/4}$, the node is sparse and \cref{thm:sparse} applies; otherwise it is dense, and \cref{thm:dense} applies, for $\lg n\ge512$.
In both cases the cost is $O(n)$.
\end{proof}

\section{Concluding remarks}\label{sec:conclusion}

Three losses contribute to the rate $1/\log n$: the up to $2$ bits wasted by the Knuth--Yao sampler at each block of entropy $B\approx\frac18\lg N$, the loss of at most $2(g+1)2^{-g}$ bits of the recycling sources at each draw, and the $O(1)$ bits wasted by each small node.
The first one is limited by the size of the tables: the decoding by table needs about $2^{4B}$ entries, so that $B$ cannot exceed a constant times $\lg N$.
Whether a relative excess $o(1/\log n)$ is compatible with $O(n)$ bit operations in the bit-RAM is the main question left open by this work.
Recycling the remainder of the Knuth--Yao walk as well, as can be done for general distributions~\cite{DraperSaad2026}, and letting $g$ grow, would remove most of these losses, but the type probabilities of a block have about $S\lg N$ bits, and we do not know how to recycle them in constant time per letter.

We have not tried to optimise the constants, which the crude thresholds of our analysis (the threshold $n^{1/4}$ for small nodes, the choice $B=\lfloor\lg N/8\rfloor$, the factor $4$ in the definition of typical types) make very large; a finer analysis, or a tilted proposal for \textsc{MergeShuffle} that makes the overflow of the majority letter much more likely, would improve them.

Two further questions concern the cost in time.
For alphabets that grow with $n$, the assembly costs $O(n(1+h)\lceil\lg(r+1)\rceil)$ bit operations, where $h$ is the entropy of the letter frequencies, and we do not know whether this can be reduced to $O(n(1+\lg(r+1)))$.
Finally, enumerative decoding reads at most $\Ent+2$ bits, and its exact bit complexity in the bit-RAM would be worth determining, as it is the natural point of comparison for samplers of this kind.

\end{document}